\documentclass[aps,twocolumn,footinbib,
letterpaper,showpacs]{revtex4}
\usepackage{graphicx}\usepackage{natbib}\usepackage{amsmath}\usepackage{times}
\usepackage{epsfig}
\usepackage{hyperref}

\def\tst{\tilde t}
\def\ttau{\tilde \tau}

\begin{document}

\title{CMSSM Spectroscopy in light of PAMELA and ATIC}

\author{Ilia Gogoladze\footnote{On leave of absence from: Andronikashvili Institute of Physics, GAS, \\ Tbilisi, Georgia.}}
\email{ilia@bartol.udel.edu}
\affiliation{Bartol Research Institute, Department of Physics and Astronomy\\
University of Delaware, Newark, Delaware 19716}

\author{Rizwan Khalid}
\email{rizwan@udel.edu}
\affiliation{Bartol Research Institute, Department of Physics and Astronomy\\
University of Delaware, Newark, Delaware 19716}

\author{Qaisar Shafi}
\email{shafi@bartol.udel.edu}
\affiliation{Bartol Research Institute, Department of Physics and Astronomy\\
University of Delaware, Newark, Delaware 19716}

\author{Hasan Y{\"u}ksel}
\email{yuksel@udel.edu}
\affiliation{Bartol Research Institute, Department of Physics and Astronomy\\
University of Delaware, Newark, Delaware 19716}

\date{January 20, 2008}

\begin{abstract}

Dark matter neutralinos in the constrained minimal supersymmetric model (CMSSM) may account for the recent cosmic ray electron and positron observations reported by the PAMELA and ATIC experiments either through self annihilation or via decay. 
However, to achieve this, both scenarios require  new physics beyond the `standard' CMSSM, and a unified explanation of the two experiments suggests  a neutralino mass of order 700 GeV - 2 TeV.  
A relatively light neutralino with mass around 100 GeV (300 GeV) can accomodate the PAMELA but not the ATIC observations based on a model of annihilating (decaying) neutralinos. 
We study the implications of these scenarios for Higgs and sparticle spectroscopy in the CMSSM and highlight some benchmark points.
An estimate of neutrino flux expected from the annihilating and decaying neutralino scenarios is provided.

\end{abstract}

\pacs{95.35.+d, 98.62.Gq, 98.70.Vc, 95.85.Ry}

\maketitle

\section{Introduction}

It is generally accepted that nearly 23\% of the universe's energy density resides in the form of non-luminous `dark' matter~\cite{Komatsu:2008hk}. This is a new form of matter which is non-baryonic and manifests itself primarily through its gravitational interactions. The highly successful Standard Model (SM) of strong, weak and electromagnetic interactions does not possess a viable dark matter candidate. Thus, new physics beyond the SM is required to incorporate dark matter, and many potential dark matter candidates have been proposed in the literature \cite{Jungman:1995df}.

Supersymmetry, more precisely MSSM, (minimal supersymmetric SM), with R-parity conservation, is arguably the most compelling extension of the SM. The MSSM predicts the existence of a stable new elementary particle called the neutralino (lightest supersymmetric particle).  With mass of order 100 GeV -- TeV, the thermal relic abundance of the lightest neutralino has the right order of magnitude to account for the observed dark matter density.

Many recent investigations of the MSSM have focused on a theoretically well motivated special case called the CMSSM \cite{Feldman:2008hs} (constrained MSSM, based on supergravity) which is far more predictive than the generic MSSM version. The latter can have more than a hundred free parameters, in contrast to the CMSSM with just 5 or so parameters. In these investigations the stable neutralino is usually found to be relatively light (100--few hundred GeV), and its indirect discovery relies on detecting cosmic ray signals (including positrons, antiprotons, gamma rays, etc.) from neutralino decay or pair annihilation in the galactic halo, galactic center, and the haloes of nearby galaxies.

The PAMELA experiment is currently taking data of  high energy anti-proton and positron fluxes, and their most recent publication claims a significant positron `excess'~\cite{Adriani:2008zr} with no corresponding anti-proton excess~\cite{Adriani:2008zq}. This result appears to confirm previous results from HEAT \cite{Barwick:1997ig} and AMS \cite{Aguilar:2007yf} within the error bars. It has been pointed out that pulsars and/or other nearby astrophysical sources may account for the PAMELA results~\cite{Aharonian(1995),Yuksel:2008rf}.
More recently, the ATIC experiment ~\cite{chang:2008zz} (see also PPB-BETS~\cite{Torii:2008xu}) has reported an appreciable flux of electrons and positrons at energies around 100 -- 800 GeV, which appears to be considerably higher than the expected background at these energies. 

A unified explanation of the PAMELA and ATIC experiments involving neutralino as the dark matter candidate could be based on one of the following two mechanisms:

\begin{itemize}
\item The dark matter is a stable neutralino with mass around 700 GeV which primarily annihilates into leptons through new interactions which lie outside the MSSM framework \cite{dm}. Depending on the framework chosen, this scenario also invokes some `boost' factor physics such as Sommerfeld enhancement~\cite{sommerfeld}.
\item The dark matter is not entirely stable \cite{Yin:2008bs, Ishiwata:2008cv} but extremely long-lived, with a lifetime $\sim 10^{26}$ sec. For neutralino dark matter,  one could introduce suitably `tiny' R (or `matter') -- parity violating couplings which satisfy the lifetime constraint and allow the neutralino to decay primarily into leptons. The tiny ($\sim 10^{-13}$) R-parity violating couplings can be understood through non-renormalizable couplings with additional discrete symmetries~\cite{Arvanitaki:2008hq}. A simple example of this is provided by the R-parity violating superpotential coupling $L L E^c$, which leads to a three-body decay mode for the neutralino~\cite{Yin:2008bs}. With a neutralino mass of around 2 TeV, this can simultaneously explain the PAMELA and  ATIC data.
\end{itemize}

If one ignores say the ATIC result, it is possible to explain the PAMELA observations with a decaying neutralino of mass $\sim 300$ GeV, which would make supersymmetry, and in particular the sparticles, far more accessible at the LHC. Even lighter ($\sim100~{\rm GeV}$) neutralino mass is feasible in the annihilation scenario.

Motivated by the  PAMELA and ATIC observations we have performed an ISAJET \cite{ISAJET} based analysis of CMSSM spectrocopy in which particular attention is paid to those regions of the CMSSM parameter space which contain heavy ($\sim 300$ GeV -- 2 TeV) neutralinos, with a relic abundance consistent with the WMAP 5 dark matter bounds~\cite{Komatsu:2008hk}, and which provide a unified explanation of the PAMELA and ATIC observations. 

The plan of the paper is as follows. We summarize the observations by PAMELA/ATIC experiments in Section II and review their implications in the context of dark matter decays or annihilations in Section III. In Section IV an estimate of neutrino flux expected from the annihilating and decaying neutralino scenarios is provided. Section V discusses how the CMSSM could be supplemented by new physics while preserving the neutralino cold dark matter framework.  We outline in Section~\ref{constraints_section} the procedure that we use to scan the CMSSM parameter space and the various experimental bounds that we take into account. In Section~\ref{results} we present plots displaying the relevant CMSSM parameter space and highlight in Tables \ref{table1} and \ref{table2} a few benchmark points that are consistent with the PAMELA and ATIC measurements.  Our conclusions are summarized in Section~\ref{conclusions}.

\section{{Observations}}

The spectra of cosmic ray electrons and positrons should have contributions from known sources including $e^-$ accelerated in supernova remnants and $e^\pm$ from collisions between cosmic rays and interstellar protons~\cite{Stecker:1970sw}.  Besides well known guaranteed backgrounds, any evidence for an additional component may carry indications of a new phenomon.  Recently, the PAMELA~\cite{Adriani:2008zr} and ATIC~\cite{chang:2008zz} experiments have observed cosmic ray positrons and/or electrons within the energy range from several GeV up to a few TeV, and provided new hints for the long suspected excess in their fluxes~\cite{Barwick:1997ig,Aguilar:2007yf} with respect to the model expectations~\cite{Moskalenko:1997gh}.  The revitalized interest in this long standing puzzle has recently generated much interest in finding both astrophysical and perhaps more exotic explanations.

The ATIC experiment has measured the combined electron and positron flux from $\sim$30~GeV reaching energies up to several TeV as shown in Fig.~1.  The primary electron spectrum is based on calculations of Ref.~\cite{Moskalenko:1997gh} (light shaded) is also shown.  As the normalization of the primary electron spectrum is already quite uncertain, we renormalize this down by a factor $\sim 0.8$ in order to match the low energy part of the ATIC data. Unlike electrons, while there are no primary sources of positrons in this model (see also~\cite{Delahaye:2008ua}), the expected secondary positron spectrum mainly due to cosmic ray interactions (dark shaded) is also presented. There is also a much smaller secondary electron component (not shown in the figure).

The PAMELA data is originally reported as the positron fraction of total electron/positron flux up to $\sim 100$~GeV~\cite{Adriani:2008zr}.  In order to show both electron and positron data on the same figure, we have converted the reported ratios into absolute flux units using the primary electron spectrum.  As shown in the figure, the corresponding positron flux based on the PAMELA data suggests a strong deviation from the secondary positron background.  While the measurements by PAMELA below $\sim$10~GeV might depend on solar modulation effects, the excess in the positron flux measurements with respect to the model expectations is evident at higher energies.  

\begin{figure}[b!]
\includegraphics[width=\columnwidth,clip=true]{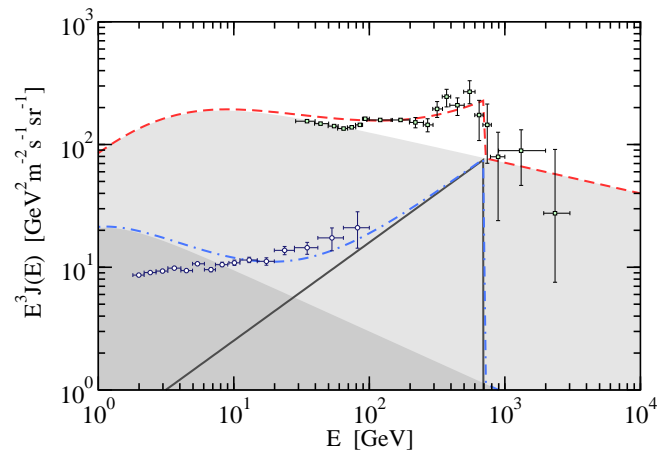}
\caption{Cosmic ray electron (light shaded) and positron (dark shaded) spectra from the model of Ref.~\cite{Moskalenko:1997gh}. Data has been taken from Ref.~\cite{chang:2008zz} (squares), and  Ref.~\cite{Adriani:2008zr} (circles).  An additional contribution of $e^\pm$ pairs with spectra $J(E)\simeq 0.4 \, (E/{\rm GeV})^{-2.2}$ GeV$^{-1} $m$^{-2} $s$^{-1} $sr$^{-1}$ with a cut-off around $\simeq 700$~GeV (solid line) can provide a better fit to both ATIC (dashed lined) and PAMELA (dot dashed line) measurements.
\label{fig:1}}
\end{figure}

\begin{figure*}[t]
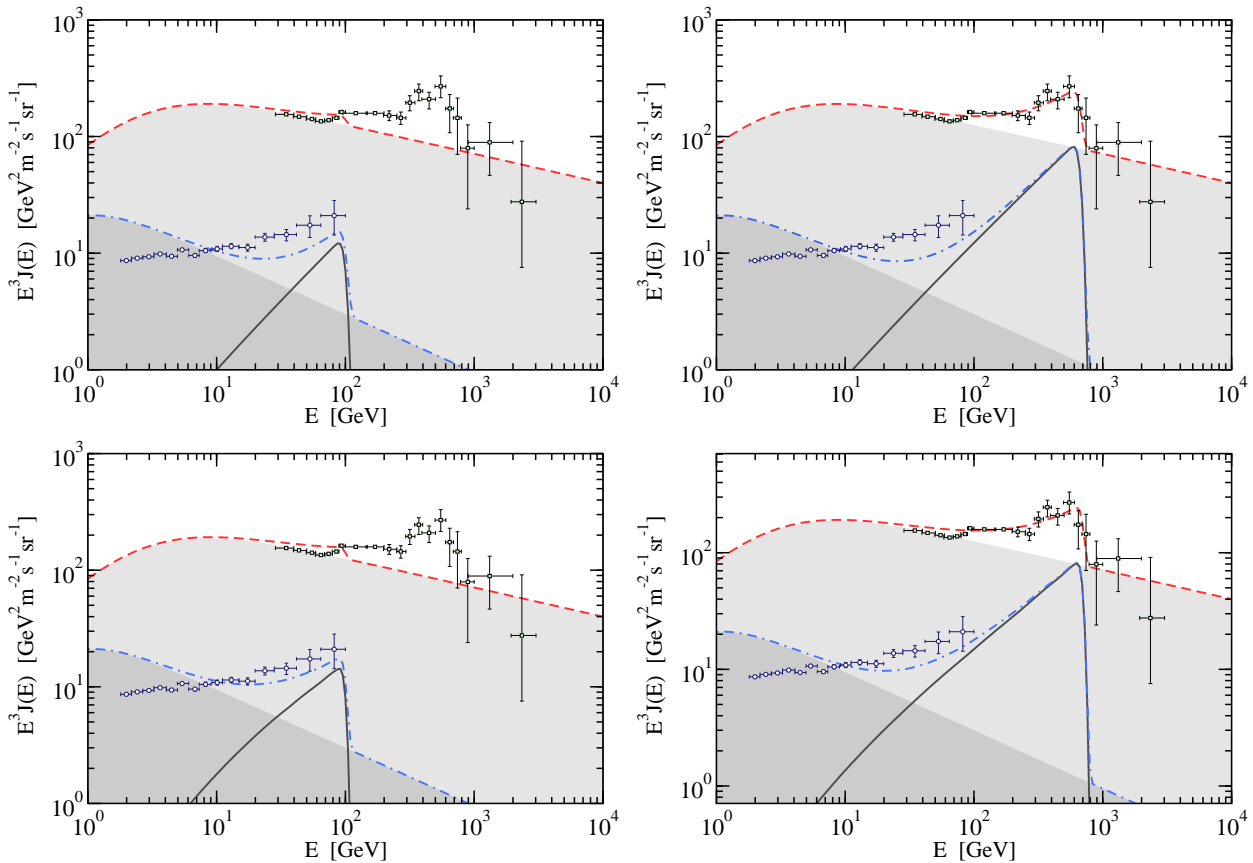

\centering
\begin{tabular}{cc}
\includegraphics[width=0.95\columnwidth,clip=true]{spectra2} &
\includegraphics[width=0.95\columnwidth,clip=true]{spectra3} \\
\includegraphics[width=0.95\columnwidth,clip=true]{spectra4} &
\includegraphics[width=0.95\columnwidth,clip=true]{spectra5}
\end{tabular}
\caption{
Models of cosmic ray electron/positron background spectra and observations as in Fig.~\ref{fig:1}.  Here the additional $e^\pm$ contribution (solid line) is assumed to be originated from dark matter decay (top panels) or annihilation (bottom panels) and aimed at explaining either only PAMELA (left panels; as a lower bound on dark matter mass) or also ATIC (right panels; as an upper bound on dark matter mass). See text for details.
\label{fig:2}}
\end{figure*}

The ATIC measurement of the total electron/positron flux also shows some excess beyond $\sim$100~GeV, most prominently at $\sim700$~GeV (once the baseline for the primary electron spectrum is normalized to the ATIC data at low energies where the positron contribution is small).  Even though the reported excess in the ATIC data is not large compared to their uncertainties~\cite{chang:2008zz} and possible systematic effects, and earlier experiments in a similar energy range show a rather large scatter (see e.g. compilation in~\cite{Torii:2008xu}), ATIC observations nonetheless provide avenues for exploring new physics, especially when taken together with PAMELA.

One can interpret the combined observations of PAMELA and ATIC as a hint of an unknown source capable of producing comparable numbers of electrons and positrons in the energy range of $10$--$700$~GeV.  For instance, if the flux of excess $e^\pm$ pairs follows, approximately, a power law of the form
\begin{equation}
J(E) \simeq 0.4 \, \left(\frac{E}{\rm GeV}\right)^{-2.2}\; {\rm GeV}^{-1} \;{\rm m}^{-2}\; {\rm s}^{-1} \; {\rm sr}^{-1} \, ,
\end{equation}
with a cut-off around $\simeq 700$~GeV (solid line in Fig.~1), then both ATIC (dashed line) and PAMELA (dot-dashed line) requirements can be reasonably satisfied. 

A very promising possibility that can account for the excess is astrophysical in origin, i.e., pulsar wind nebulae which could accelerate $e^\pm$ pairs produced around a neutron star, and which would naturally yield equal numbers of positrons and electrons.  Indeed, the tentative detection of TeV gamma rays from Geminga~\cite{Abdo:2007ad}, a nearby pulsar, immediately suggests the possibility of particle accelaration up to multi--TeV energies~\cite{Yuksel:2008rf} (provided that the origin of the gamma rays is inverse Compton scattering).  Unlike electrons, which can also be accelarated in supernova remnants, positrons have no other known primary sources, and thus the resulting excess in positron flux could be very prominent and similar to the PAMELA observations~\cite{Yuksel:2008rf,Aharonian(1995)}.

\section{{Positrons/Electrons from Dark Matter}}

Another exciting explanation is that some exotic mechanism such as dark matter decay or annihilation, which are the main subjects of this study, might be responsible for the excesses seen in observations.  We will assume throughout that the dark matter is the lightest CMSSM neutralino. For a study on the signals from dark matter annihilation/decay products, an understanding of the distribution of dark matter within the galaxy is essential.  Simulations of cold dark matter suggest that dark matter assembles in halos.  The dark matter distribution within the halo is usually taken to be spherically symmetric and the density profile can be expressed as function of the distance from the center of the halo, 
\begin{equation}
\rho(r) = \frac{\rho_0}{(r/r_s)^\gamma[1+(r/r_s)^\alpha]^{(\beta-\gamma)/\alpha}}, \, .
\label{eq:fitting}
\end{equation}
where the parameters $(\alpha, \beta, \gamma, r_s)=(1, 3, 1, {\rm 20 \; kpc})$ for the NFW profile~\cite{Navarro:1995iw}, which also suggests that the density of dark matter at the solar circle is $\rho(R_{sc} \simeq 8.5$ ~kpc~$)= \rho_{sc} \simeq 0.3$ GeV/cm$^3$.

As our main motivation is to account for the observations of ATIC and PAMELA, we shall focus on generic leptophilic scenarios in which the dominant annihilation/decay products are assumed to be positrons and electrons.  In such models, once electrons/positrons are created, their density $n$ at a given place, time and energy, is governed by the diffusion equation,
\begin{equation}
\frac{\partial n}{\partial t} =  {\mathbf \nabla} \cdot
\left[ {\mathcal D} (E) {\mathbf \nabla} n \right]  +
\frac{\partial}{\partial E} \left[ \ell(E) \, n \right] 
+ Q ( {\mathbf r} , E ) \, ,
\end{equation}
where $Q( {\mathbf r} , E )$ is the source term describing the particle injection rate as discussed below, $\ell(E)= \ell_0 (E/{\rm GeV})^{-2}$ is the energy loss rate and  ${\mathcal D} (E)= {\mathcal D}_0 (E/{\rm GeV})^\delta$ is the diffusion coefficient, both assumed to be independent of space.

The spectrum of a particle per dark matter annihilation, $ \chi \,\chi \rightarrow e^+ e^- $, can be written as $ \phi^{a}(E)=\delta(E-m_\chi) $ in terms of the Dirac $\delta$ function, where $m_\chi$ is the mass of the dark matter particle.  The corresponding source term is then
\begin{equation}
Q^{a}({\mathbf r},E) =\left[\frac{\rho({\mathbf r})}{\rho_{sc}}\right]^2
\left[ \frac{1}{2} \frac{\rho_{sc}^2}{m_\chi^2} {\rm f_B} \langle \sigma v \rangle \right] \phi^{a}(E) \, ,
\end{equation}
where the factor of $1/2$ arises by assuming that the dark matter candidate is its own antiparticle. Here  the term ${\rm f_B} \langle \sigma v \rangle$ is the product of the annihilation cross section and the relative velocity together with the requried `boost factor ${\rm f_B}$' that may come from either new particle physics or astrophysical enhancements.   

If the dark matter decays into three light particles such as $\chi \rightarrow e^+ e^- \nu $, the spectrum of each daughter particle, to a rough approximation, is given by $\phi^{d}(E) \simeq \delta(E-m_\chi/3) $ with the corresponding source term
\begin{equation}
Q^{d}({\mathbf r},E) =\frac{\rho({\mathbf r})}{\rho_{sc}}
\left[ \frac{\rho_{sc}}{m_\chi \tau}\right] \phi^{d}(E) \, ,
\end{equation}
where $\tau$ is the lifetime corresponding to this decay channel, which may arise from new physics.  Note that the prefactor $\rho_{sc}$ is introduced to make the term in the second paranthesis dimensionless and will cancel out in our final results.

With these definitions of the source terms, we follow an approach similar to Ref.~\cite{Delahaye:2007fr,Ibarra:2008qg} in evaluating the local particle flux.  Assuming steady-state conditions, typical in assessing the particle flux from dark matter annihilation/decay, the solution of the diffusion equation can be cast as
\begin{equation}
n(E)=\kappa\int_{E}^{\infty} dE^{'} \phi(E^{'}) \zeta(E,E^{'}) \, .
\label{eq:difsol}
\end{equation}
The coefficient $\kappa$ encodes particle physics input such that for annihilations $\kappa^a= \frac{1}{2} ({\rho_{sc}^2}/ {m_\chi^2}) {\rm f_B} \langle \sigma v \rangle$, while for decays $\kappa^d= {\rho_{sc}} /{(m_\chi \tau)}$. The spectrum of each given daughter particle per annhilation or decay is $\phi$.  The function ${\cal \zeta}$ encodes dependence of the solution on the chosen dark matter halo profile as well as particle propagation and diffusion and has a form similar to $\tilde{I}$ in Ref.~\cite{Delahaye:2007fr} for annihilations up to multiplicative factors.  The flux of particles is simply $J(E)=({c}/{4\pi}) n(E)$, where $c$ is the speed of light.

The meaning of Eqn.~\ref{eq:difsol} is as follows. For a particle to be observed locally at a given energy $E$, it has to be produced at a higher energy $E^{'}$ at a more distant location in the galaxy. Then the particle will lose energy and diffuse until it reaches the solar system, abiding by the constraints of the diffusion equation.  As high energy particles loose energy fast, their observed flux should be dominated by nearby processes within a few kiloparsecs.  Higher production rate for rather cuspy halo profiles, especially for annihilations  around the central parts of the galaxy are only important at lower energies.  Since we focus on leptophilic scenarios, plausible severe constraints due to final states other than $e^\pm$ pairs, such as antiprotons, are already avoided.  Thus the exact parametrization of the underlying dark matter halo profile is less relevant as long as the local normalization of dark matter density, $\rho_{sc}\simeq 0.3$ GeV/cm$^3$ is consistent.

Besides electrons/positrons, if other charged leptonic intermediate states exist, such as tau that decay to $e^\pm$, the fit of the theoretical expectations to the observations might be visually improved.  However, an exact reproduction of the observations cannot be essential since (1) the observations are highly uncertain, and more importantly (2) the background fluxes are not well known and already strongly dependent on models.  Since an attempt to over--constrain the theories while assuming a perfect knowledge of the background electron/positron fluxes may not yield additional useful results, we choose to probe only the most generic features of the observations by PAMELA and ATIC and present our results for a NFW dark matter profile together with the MED diffusion model~~\cite{Delahaye:2007fr} with appropriate parameters $(D_0=0.0112 \; {\rm pc}^2/{\rm yr},\delta=0.7,\ell_0=10^{-16}\;{\rm GeV}/{\rm s})$.

Next we outline how the PAMELA and ATIC observations could be accounted for by a neutralino dark matter scenario through annihilation or decay, and what would be the expected ranges of parameters such as dark matter mass, lifetime or annihilation cross section under these assumptions.  We identify two scenarios which can reasonably account for the positron excess observed by PAMELA with a relatively light ($\sim 100-300~{\rm GeV}$) dark matter candidate. In addition we consider two scenarios with an appropriately heavy ($\sim 700~{\rm GeV}-2~{\rm TeV}$) dark matter candidate which provide a unified explanation of both the PAMELA and ATIC observations.

In Fig.~2, the additional $e^\pm$ contribution (solid line) is assumed to originate from dark matter decay or annihilation.  The dashed line shows the expectation for total electron and positron flux when both e$^-$ and e$^+$ fluxes are added to the primary electron background (light shaded).  This can be compared to the ATIC data which corresponds to the total lepton spectrum.  We also show the total positron spectrum (dot-dashed line) by adding only the additional positron flux from dark matter annihilation or decay to the secondary positron spectrum (dark shaded), which is then compared to the PAMELA data.  The theoretical fits are slightly smoothed out to simulate finite resolution of experiments. The four panels displayed are:

\begin{itemize}
\item{\it Top-Left:} A decaying dark matter scenario with a lifetime $\tau \sim 2 \times 10^{27}$~s and $m_\chi \sim 0.3$~TeV which is the minimal mass required to account for PAMELA (dot-dashed line) without causing significant tension with ATIC (dashed-line).
\item{\it Top-Right:} A decaying dark matter scenario with a lifetime $\tau \sim 3 \times 10^{26}$~s and $m_\chi \sim 2$~TeV which is the maximum mass that can account for ATIC observation with rather sharp cut-off around $\sim 0.7$ TeV (dashed line) and also account for PAMELA (dot-dashed line). Thus, masses in the range of $\sim 0.3$--$2$ TeV with a lifetime of $\sim 10^{27}$~s can be regarded as acceptable for a decaying neutralino dark matter scenario.
\item{\it Bottom-Left:} An annihilating dark matter scenario with ${\rm f_B} \langle \sigma v\rangle \sim 1.5\times 10^{-25}$ cm$^3$/s and $m_\chi \sim 0.1$~TeV, which corresponds to the minimal mass required to explain PAMELA (dot-dashed line) without significantly exceeding the ATIC measurement (dashed-line). 
\item{\it Bottom-Right:}  An annihilating dark matter scenario with ${\rm f_B} \langle \sigma v\rangle \sim 6\times 10^{-24}$ cm$^3$/s and $m_\chi \sim 0.7$~TeV that can simultaneously fit the PAMELA excess (dot-dashed line) and ATIC excess and also the observation of rather sharp cut-off around $\sim 0.7$ TeV (dashed line).  For an annihilating scenario, the preferred mass ranges are $\sim 0.1$--$0.7$ TeV with ${\rm f_B} \langle \sigma v\rangle \sim 10^{-24}$ cm$^3$/s.
\end{itemize}

The predictions for dark matter annihilation and decay barely differ only at low energies as can be seen in Fig.~2, where the flux from the former is larger.  This is due to the fact that the annihilation rate is proportional to the square of the dark matter density and any enhancement towards the Galactic Center will cause an enhanced contributions at lower energies.

\section{Neutrinos from Dark Matter}

\begin{figure}[b!]
\includegraphics[width=\columnwidth,clip=true]{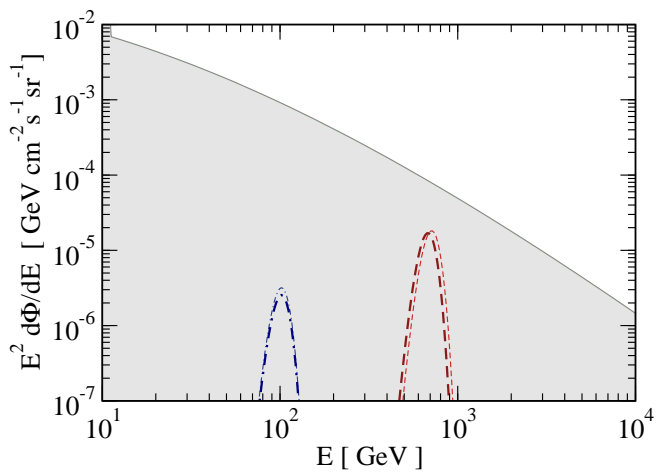}
\caption{The atmospheric neutrino spectrum averaged over whole sky (solid line) is compared  to the total neutrino flux associated with the parameters used in { top-left} ({ top-right}, {bottom-left}, { bottom-right}) panel of Fig.~2 which is denoted with a thin-dot-dashed (thin-dashed, thick-dot-dashed, thick-dashed) line.  See text for details.
\label{fig:3}}
\end{figure}

An interesting application of these results is the plausible presence of other leptonic final states such as neutrinos. Dark matter annihilating dominantly into neutrinos has been considered in Ref.~\cite{Beacom:2006tt,Yuksel:2007ac} and more recently in the context of more recent observations~\cite{Hisano:2008ah,Liu:2008ci}. For instance, neutralino decay may  yield neutrinos besides pairs of $e^{\pm}$.  Similarly, it is conceivable that in annihilation, neutrinos and $e^{\pm}$ pairs are produced in comparable numbers.  

To further illustrate the implication of the four scenarios, we calculate the total neutrino flux (utilizing prescriptions outlined in Refs.~\cite{Yuksel:2007ac,Yuksel:2007xh}) using the above parameters and present the results in Fig.~3. The total neutrino flux for these bumps is $\sim 6\times 10^{-9}$ cm$^{-2} $s$^{-1} $sr$^{-1}$ when averaged over the whole sky (both smoothed with a 10\% Gaussian in the figure).  The fluxes from the decay (thick lines) and annihilation (thin lines) scenarios are comparable as their main normalizations are provided by the requirement to account for the PAMELA and ATIC observations.  The fluxes are also averaged over the whole sky which dampens any relative enhancement of annihilation compared to decay scenarios, especially towards the center of the Galaxy. 

These can be compared to the atmospheric neutrino spectrum (solid line) which is based on measurements by Refs.~\cite{Desai:2004pq,Daum:1994bf,Achterberg:2007qp,Achterberg:2005fs,Ashie:2005ik} and also agrees with the theoretical modelling of Ref.~\cite{Gaisser:2002jj}.  The bumps at $\sim 100$ GeV which are devised to account for PAMELA only fall short by several orders of magnitude below the atmospheric neutrino flux. However at energies $\sim 700$ GeV, the corresponding neutrino fluxes are within reach of the atmospheric neutrino background, thus offer hope to test this scenario in the near future.

As we have focused here on leptophilic scenarios, we do not discuss models in which gamma ray fluxes are among the primary decay or annihilation products (see e.g. Refs.~\cite{Mack:2008wu,Yuksel:2007dr} where the dominant dark matter annihilation/decay products are gamma rays). However, gamma rays might accompany the production of $e^\pm$ pairs such as through internal bremsstrahlung~\cite{Bell:2008vx}.  Also inverse Compton and synchrotron emission might be produced from the electrons and positrons.  In such cases, especially for annihilations, dark matter profiles that are shallower around the center of the Galaxy compared to the NFW parametrization might be preferred~\cite{Cholis:2008wq}.

\begin{figure*}
\centering
\includegraphics{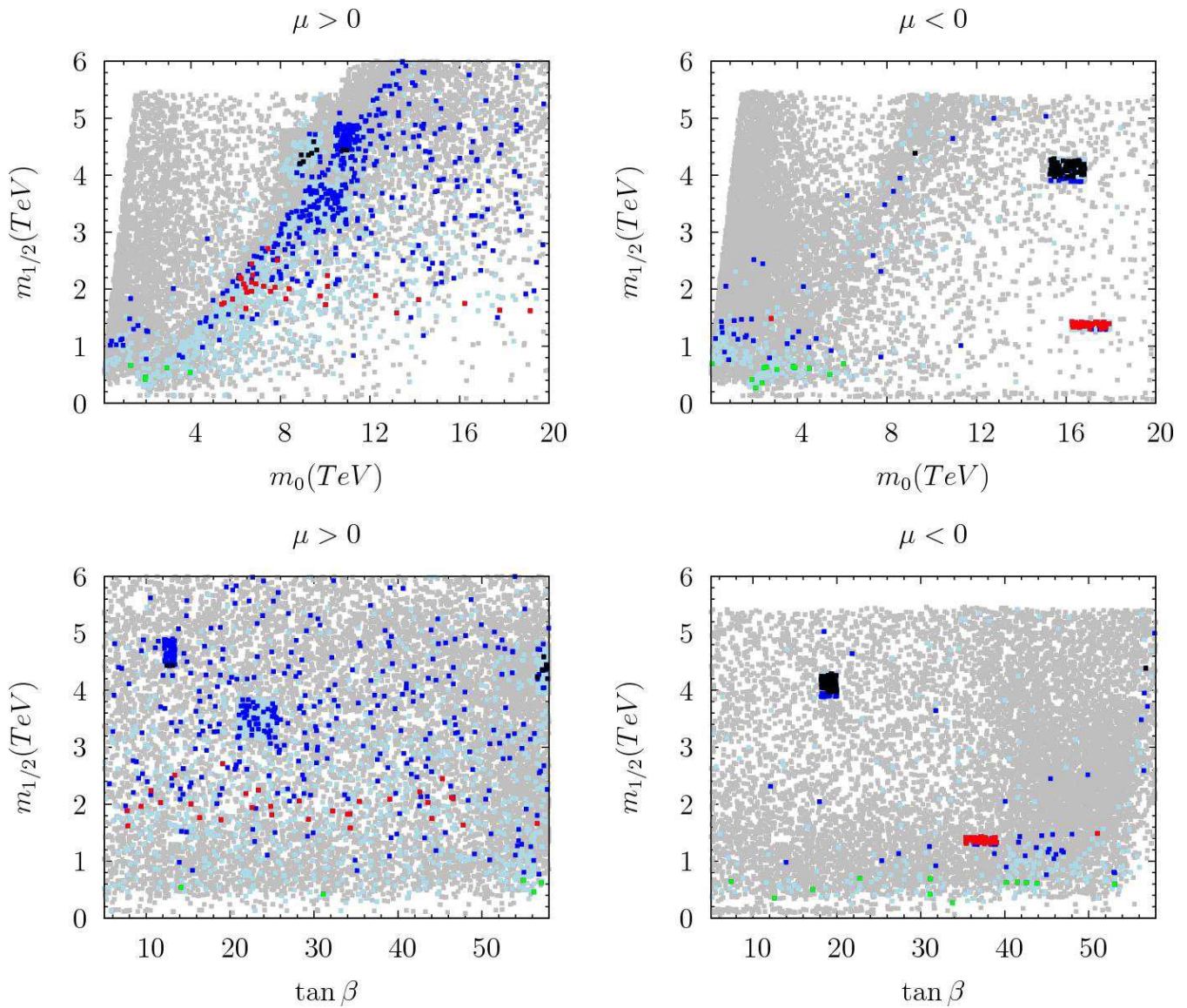}
\caption{Plots in ($m_{1/2}$, $m_{0}$) and ($m_{1/2}$, $\tan\beta$) planes for ${\mu}>0$ (left panel) and ${\mu}<0$ (right panel). Gray points satisfy constraints from colliders ( $BR(B_s\rightarrow \mu^+ \mu^-)$, $BR(B\rightarrow X_{s} \gamma)$, and the chargino and Higgs mass bounds). Light blue points satisfy the WMAP upper bound on dark matter relic abundance. Green ($m_{\tilde{\chi}^0_1} \le 0.3~{\rm TeV}$), red ($0.65~{\rm TeV} \le m_{\tilde{\chi}^0_1} \le 0.75~{\rm TeV}$),  black ($1.9~{\rm TeV} \le m_{\tilde{\chi}^0_1} \le 2.1~{\rm TeV}$), and dark blue ($m_{\tilde{\chi}^0_1} \le 2.5~{\rm TeV}$) points satisfy both the upper and lower bounds on dark matter relic abundance. Green, red and black points are subsets of dark blue points. Dense regions in all figures correspond to parameter values that are especially relevant for PAMELA and ATIC, and for which additional data has been accumulated. \label{fund1}}
\end{figure*}

\begin{figure*}
\centering
\includegraphics{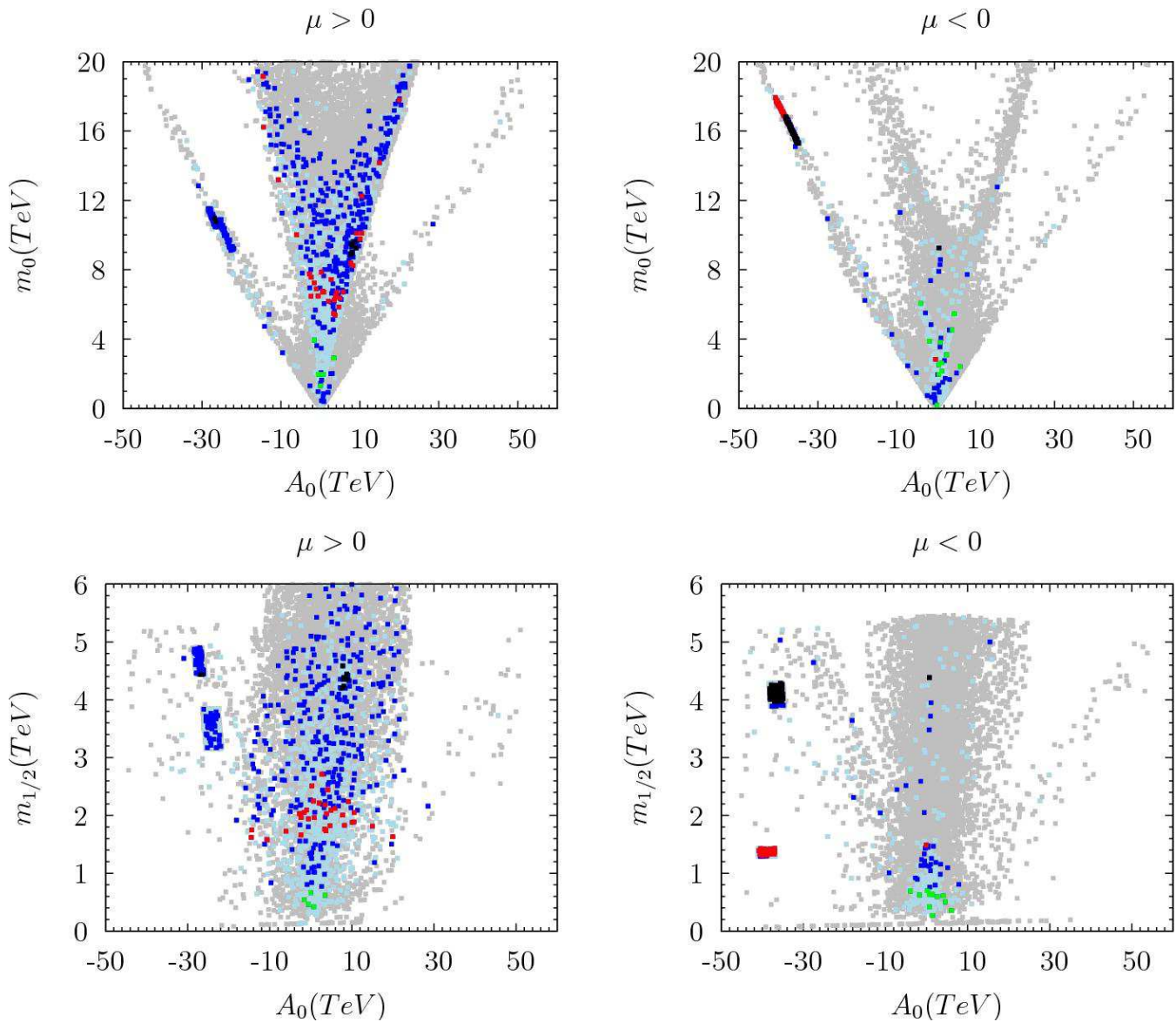}
\caption{Plots in ($m_{0}$, $A_{0}$) and ($m_{1/2}$, $A_0$) planes for $\mu>0$ (left panel) and $\mu<0$ (right panel). Color coding same as in Fig. \ref{fund1}.\label{fund2}}
\end{figure*}

\begin{figure*}
\centering
\includegraphics{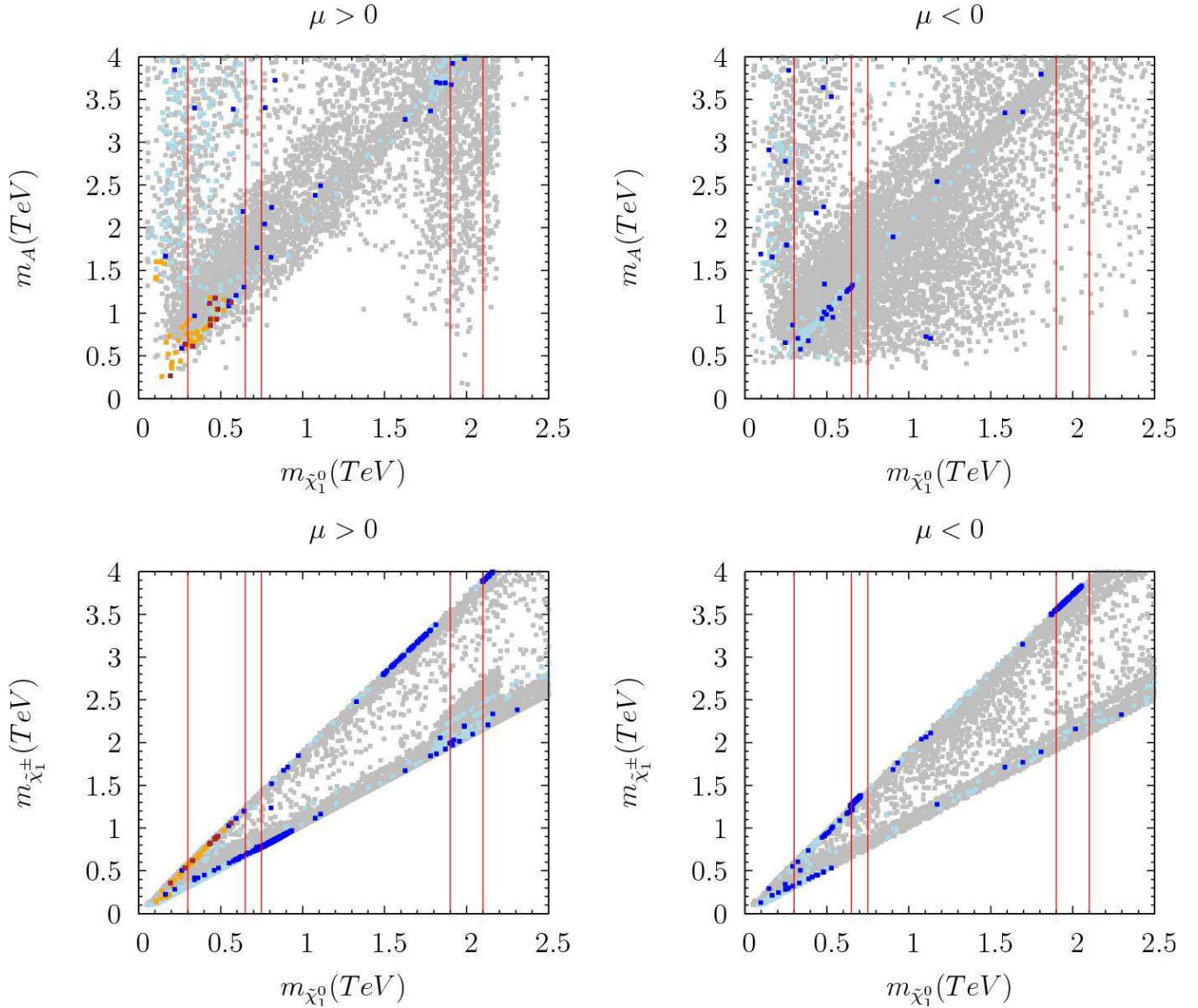}
\caption{Plots in ($m_A$, $m_{\tilde{\chi}^0_1}$) and ($m_{\tilde{\chi}^{\pm}_{1}}$, $m_{\tilde{\chi}^0_1}$) planes for $\mu>0$ (left panel) and $\mu<0$ (right panel). Gray points satisfy constraints from colliders ( $BR(B_s\rightarrow \mu^+ \mu^-)$, $BR(B\rightarrow X_{s} \gamma)$, and the chargino and Higgs mass bounds). Light blue points satisfy the WMAP upper bound on dark matter relic abundance. Dark blue points satisfy both the upper and lower bounds on dark matter relic abundance. Orange and brown points satisfy the constraint from $\Delta a_\mu$ (for $\mu>0$, as expected). Orange points satisfy only the lower bound on dark matter relic density, while brown ones satisfy both the upper and lower bounds. Vertical lines correspond to $m_{\tilde{\chi}^0_1}=0.3,0.65,0.75,1.9~{\rm and}~2.1~{\rm TeV}$. ATIC imposes $0.65~{\rm TeV}\le m_{\tilde{\chi}^0_1}\le0.75~{\rm TeV}$ ($1.9~{\rm TeV}\le m_{\tilde{\chi}^0_1}\le2.1~{\rm TeV}$) based on annihilating (decaying) neutralinos. PAMELA requires $m_{\tilde{\chi}^0_1}\ge0.1~{\rm TeV}$ ($m_{\tilde{\chi}^0_1}\ge0.3~{\rm TeV}$) based on annihilating (decaying) neutralinos. \label{spar1}}
\end{figure*}

\begin{figure*}
\centering
\includegraphics{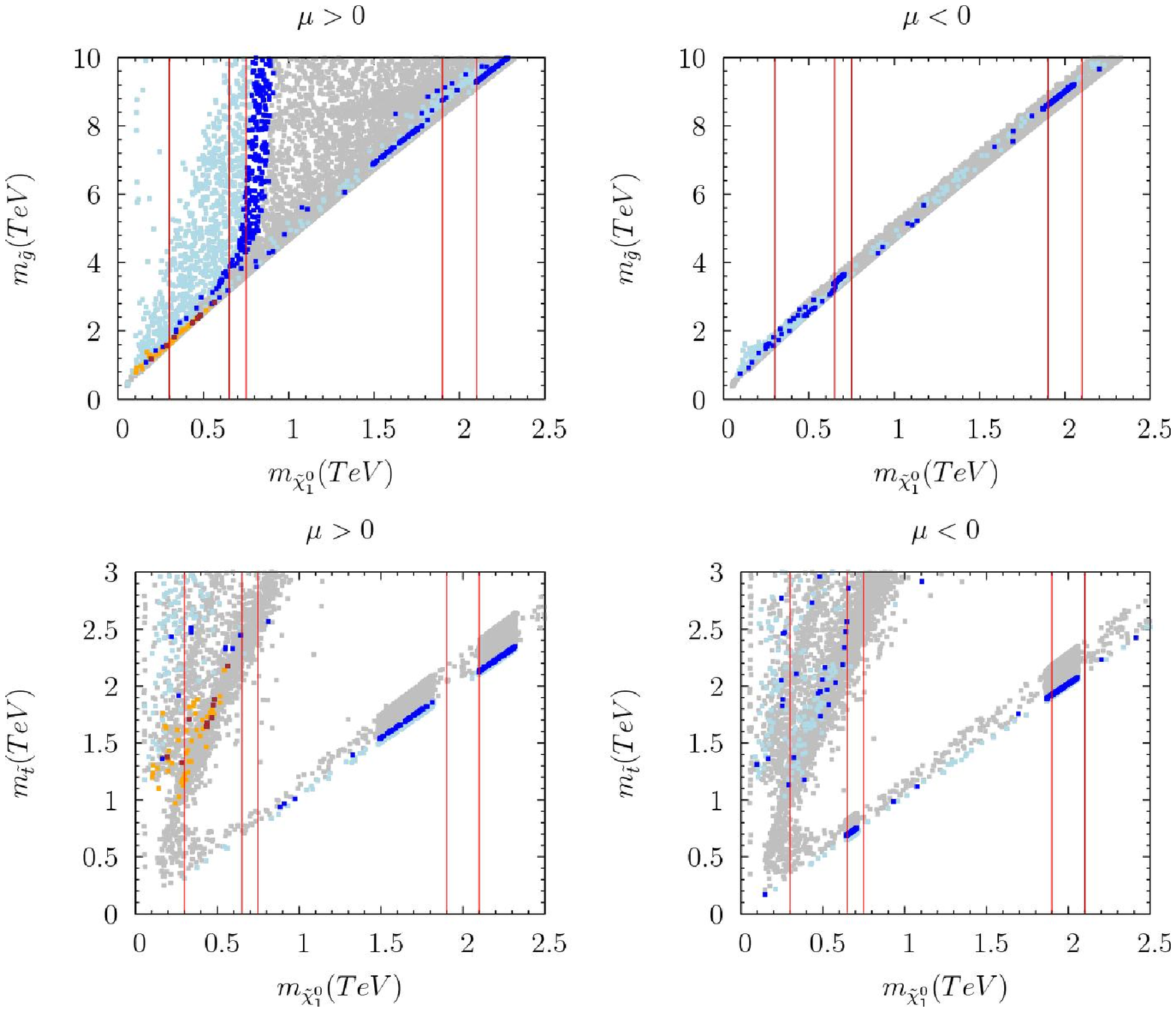}
\caption{Plots in ($m_{\tilde g}$, $m_{\tilde{\chi}^0_1}$) and ($m_{\tilde t}$, $m_{\tilde{\chi}^0_1}$) planes for $\mu>0$ (left panel) and $\mu<0$ (right panel). Color coding same as in Fig. \ref{spar1}. Vertical lines correspond to $m_{\tilde{\chi}^0_1}=0.3,0.65,0.75,1.9~{\rm and}~2.1~{\rm TeV}$. ATIC imposes $0.65~{\rm TeV}\le m_{\tilde{\chi}^0_1}\le0.75~{\rm TeV}$ ($1.9~{\rm TeV}\le m_{\tilde{\chi}^0_1}\le2.1~{\rm TeV}$) based on annihilating (decaying) neutralinos. PAMELA requires $m_{\tilde{\chi}^0_1}\ge0.1~{\rm TeV}$ ($m_{\tilde{\chi}^0_1}\ge0.3~{\rm TeV}$) based on annihilating (decaying) neutralinos. \label{spar2}}
\end{figure*}

\begin{figure*}
\centering
\includegraphics{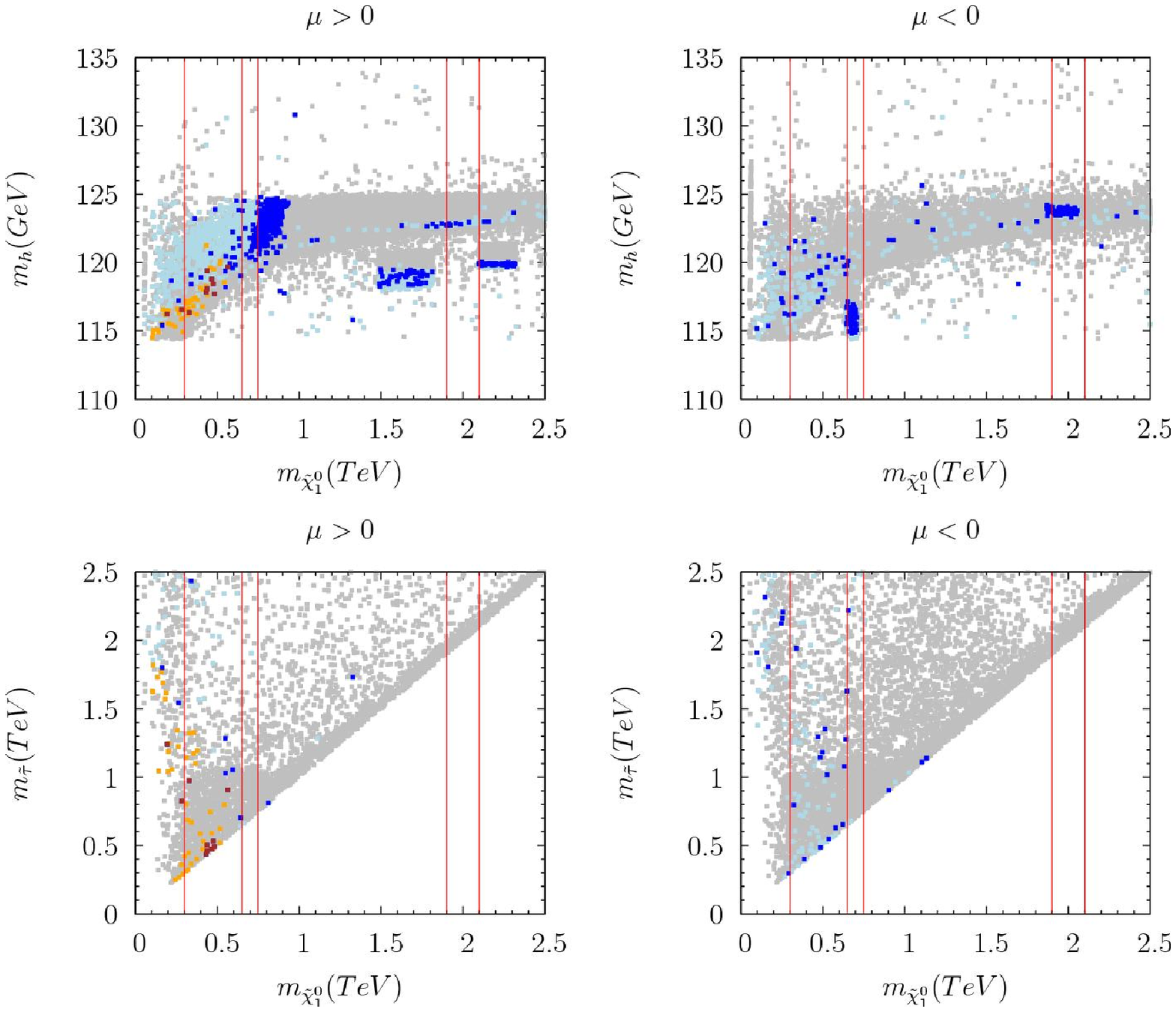}
\caption{Plots in ($m_{h}$, $m_{\tilde{\chi}^0_1}$) and ($m_{\tilde \tau}$, $m_{\tilde{\chi}^0_1}$) planes for $\mu>0$ (left panel) and $\mu<0$ (right panel). Color coding same as in Fig. \ref{spar1}. Vertical lines correspond to $m_{\tilde{\chi}^0_1}=0.3,0.65,0.75,1.9~{\rm and}~2.1~{\rm TeV}$. ATIC imposes $0.65~{\rm TeV}\le m_{\tilde{\chi}^0_1}\le0.75~{\rm TeV}$ ($1.9~{\rm TeV}\le m_{\tilde{\chi}^0_1}\le2.1~{\rm TeV}$) based on the model of annihilating (decaying) neutralinos. PAMELA requires $m_{\tilde{\chi}^0_1}\ge0.1~{\rm TeV}$ ($m_{\tilde{\chi}^0_1}\ge0.3~{\rm TeV}$) based on annihilating (decaying) neutralinos. \label{spar3}}
\end{figure*}

\begin{table*}[t]
\centering
\begin{ruledtabular}
\begin{tabular}{lccc}
          & Point 1 & Point 2 & Point 3    \\
\hline
$m_{1/2}$ & 665.1   & 2051.8   & 4437.6    \\
$m_{0} $ &  1305.5  & 7251.8   & 10970    \\
$\tan\beta$ & 54.928   & 28.236 & 12.49 \\
$A_0$ &  6.975  & -1606.2 & -26842 \\
${\rm sgn}~\mu$   & +   &  +  & + \\
\hline
$m_h$          & 117   & 122   & 120    \\
$m_H$          & 642   & 6545  & 15510   \\
$m_A$          & 637   & 6502  & 15408    \\
$m_{H^{\pm}}$  & 648   & 6546  & 15536     \\
\hline
$m_{\tilde{\chi}^{\pm}_{1,2}}$
& 533, 730 & 771, 1720 & 3888, 11161  \\
$m_{\tilde{\chi}^0_{1,2,3,4}}$ & {\bf 283}, 532, 714, 730
& {\bf 743}, 753, 928, 1753
& {\bf 2099}, 3857, 11131, 11131  \\
$m_{\tilde{g}}$ & 1563 & 4638 & 9293   \\
\hline $m_{{\tilde{u}}_{L,R}}$
& 1876, 1845 & 8134, 8082 & 11311, 11080  \\
$m_{\tilde{t}_{1,2}}$
& 1327, 1537 & 5357, 6776 & 2132, 9570 \\
\hline $m_{{\tilde{d}}_{L,R}}$
& 1878, 1841 & 8135, 8076 & 13433, 13231  \\
$m_{\tilde{b}_{1,2}}$
& 1496, 1571 & 6751, 7678 & 9807, 12760  \\
\hline
$m_{\tilde{\nu}_{1,2,3}}$
& 1374, 1374, 1165 & 7363, 7363, 7129  & 11328, 11328, 11121   \\
\hline
$m_{{\tilde{e}}_{L,R}}$
& 1377, 1328  & 7362, 7284 & 11311, 11080  \\
$m_{\tilde{\tau}_{1,2}}$
& 824, 1169  & 6802, 7127 &  10648, 11114  \\
\hline
$\Omega_{CDM}h^2$ &  0.1199 & 0.1042  & 0.11375  \\
\hline
$\langle \sigma v\rangle ({\rm cm^3/s})$
& $1.495 \times 10^{-26}$
& $1.76 \times 10^{-26}$
& $6.2 \times 10^{-31}$  \\
$\sigma_{\tilde{\chi}_1^0-p, {\rm SI}}({\rm pb})$
& $1.81\times 10^{-9}$
& $4.07\times 10^{-9}$
& $3.56\times 10^{-14}$ \\
$\sigma_{\tilde{\chi}_1^0-n, {\rm SI}}({\rm pb})$
& $1.95\times 10^{-9}$
& $4.17\times 10^{-9}$
& $3.68\times 10^{-14}$ \\
\end{tabular}
\end{ruledtabular}
\caption{ Sparticle and Higgs masses (in units of GeV),
with $m_t=172.6$ GeV and $\mu>0$. We present three benchmark points. Also included are the spin-independent
neutralino-nucleon interaction cross-sections. Point 1 can explain results from PAMELA based on either a
decaying or an annihilating neutralino model. Point 2 (Point 3) can explain ATIC and PAMELA based on the model
of neutralino annihilation (decay). } \label{table1}
\end{table*}

\begin{table*}[t]
\centering
\begin{ruledtabular}
\begin{tabular}{lccc}
          & Point 1 & Point 2 & Point 3    \\
\hline
$m_{1/2}$ & 697.25   & 1307.3   & 15689    \\
$m_{0} $ &  186.39  & 17042.7   & 4161.8    \\
$\tan\beta$ & 22.73   & 37.60 & 19.65 \\
$A_0$ &  222.3  & -38866.3 & -35936 \\
${\rm sgn}~\mu$   & -   &  -  & - \\
\hline
$m_h$          & 116   & 116   & 124    \\
$m_H$          & 866   & 15075  & 19224     \\
$m_A$          & 860   & 14977  & 19099    \\
$m_{H^{\pm}}$  & 869   & 15071  & 19209    \\
\hline
$m_{\tilde{\chi}^{\pm}_{1,2}}$
& 552, 809 & 1271, 12085 & 3727, 13326  \\
$m_{\tilde{\chi}^0_{1,2,3,4}}$ & { 292}, 552, 780, 809
& {\bf 650}, 1258, 12024, 12025
& {\bf 1997}, 3692, 13290, 13290  \\
$m_{\tilde{g}}$ & 1567 & 3396 & 8988   \\
\hline $m_{{ \tilde{u}}_{L,R}}$
& 1439, 1386 & 17146, 17183 & 17230, 17131  \\
$m_{\tilde{t}_{1,2}}$
& 1132, 1354 & 694, 10027 & 2014, 11683 \\
\hline $m_{{ \tilde{d}}_{L,R}}$
& 1442, 1380 & 17146, 17190 & 17230, 17120  \\
$m_{\tilde{b}_{1,2}}$
& 1317, 1351 & 10287, 14164 & 11938, 16008  \\
\hline
$m_{\tilde{\nu}_{1,2,3}}$
& 496, 496, 487 & 17068, 17068, 14620  & 15913, 15913, 15257   \\
\hline
$m_ {{\tilde{e}}_{L,R}}$
& 506, 320  & 17045, 17040 & 15889, 15752  \\
$m_{\tilde{\tau}_{1,2}}$
& 295, 503  & 11919, 14708 &  14440, 15268  \\
\hline
$\Omega_{CDM}h^2$ &  0.0981 & 0.1011  & 0.0982  \\
\hline
$\langle \sigma v\rangle ({\rm cm^3/s})$
& $1.16 \times 10^{-28}$
& $9.52 \times 10^{-29}$
& $1.219 \times 10^{-30}$  \\
$\sigma_{\tilde{\chi}_1^0-p, {\rm SI}}({\rm pb})$
& $2.51\times 10^{-13}$
& $3.05\times 10^{-13}$
& $5.60\times 10^{-14}$ \\
$\sigma_{\tilde{\chi}_1^0-n, {\rm SI}}({\rm pb})$
& $5.73\times 10^{-13}$
& $3.08\times 10^{-13}$
& $5.65\times 10^{-14}$ \\
\end{tabular}
\end{ruledtabular}
\caption{ Sparticle and Higgs masses (in units of GeV), with $m_t=172.6$ GeV and $\mu<0$. We present three benchmark points. Also included are the spin-independent neutralino-nucleon interaction cross-sections.  Point 1 can explain results from PAMELA based on either a decaying or an annihilating neutralino model. Point 2 (Point 3) can explain ATIC and PAMELA based on the model of neutralino annihilation (decay).} \label{table2}
\end{table*}

\section{From Observations to CMSSM}

In order to understand the unexpectedly high electronic flux observed by PAMELA and ATIC the CMSSM must be supplemented by new physics. If one wishes to preserve the basic neutralino cold dark matter framework, which we propose to do in this paper, at least two options are available.

Perhaps the simplest modification one could contemplate is to include R-parity violating superpotential coupling(s) with suitably tiny dimensionless coefficient(s), such that the neutralino primarily decays into leptons, with lifetime $\sim 10^{26} {\rm sec}$. A simple example is provided by the coupling $LLE^c$~ \cite{Yin:2008bs}, which leads to 3-body leptonic decays of the neutralino. This modification has an important advantage in that it preserves the basic structure of the CMSSM, especially by leaving intact the standard calculations for estimating the relic neutralino abundance. The implications for CMSSM spectroscopy for this case which has a rather heavy neutralino ($\sim 2~{\rm TeV}$) are discussed in Section \ref{results}.

Dark matter neutralinos annihilating in the halos of galaxies provide an exciting new source for cosmic rays including positrons. To explain the PAMELA and ATIC observations in terms of neutralino annihilations, the CMSSM must be supplemented by new physics which should accomplish the following three things: allow neutralino annihilations primarily into leptonic channels, produce a boost factor of order $10^3$ or so through Sommerfeld enhancement~\cite{sommerfeld}, and last but by no means least, ensure that the relic abundance of neutralinos is consistent with WMAP bounds. Several interesting proposals for annihilating WIMPS that explain PAMELA and ATIC have been put forward~\cite{dm}. However, it is not obvious if they can be easily adapted to the CMSSM scenario we are considering in which the neutralino makes up the observed cold dark matter in the universe. Thus, even though an explicit particle physics construction for neutralino annihilation satisfying the above conditions is not yet available, we will in the following discussion present spectroscopy results which cover a wide range of neutralino masses currently favored by PAMELA and ATIC, including the 700 GeV or so value motivated from neutralino annihilation.

\section{Phenomenological constraints and scanning procedure\label{constraints_section}} \label{ch:constraints}

We employ ISAJET~7.78 package~\cite{ISAJET} to perform random scans over the parameter space. In this package, the weak scale values of gauge and third generation Yukawa couplings are evolved to $M_{\rm GUT}$ via the MSSM renormalization group equations (RGEs) in the $\overline{DR}$ regularization scheme, where $M_{\rm GUT}$ is defined to be the scale at which $g_1=g_2$. We do not enforce an exact unification of the strong coupling $g_3=g_1=g_2$ at $M_{\rm GUT}$, since a few percent deviation from unification can be assigned to unknown GUT-scale threshold corrections~\cite{Hisano:1992jj}. At $M_{\rm GUT}$, the CMSSM boundary conditions are imposed and all the SSB parameters, along with the gauge and Yukawa couplings, are evolved back to the weak scale $M_{\rm Z}$. In the evaluation of  Yukawa couplings the SUSY threshold corrections~\cite{Pierce:1996zz} are taken into account at the common scale $M_{\rm SUSY}= \sqrt{m_{\tst_L}m_{\tst_R}}$. The entire parameter set is iteratively run between $M_{\rm Z}$ and $M_{\rm GUT}$ using the full 2-loop RGEs until a stable solution is obtained. To better account for leading-log corrections, one-loop step-beta functions are adopted for gauge and Yukawa couplings, and the SSB parameters $m_i$ are extracted from RGEs at multiple scales $m_i=m_i(m_i)$.  The RGE-improved 1-loop effective potential is minimized at an optimized scale $M_{\rm SUSY}$, which effectively accounts for the leading 2-loop corrections. Full 1-loop radiative corrections are incorporated for all sparticle masses.

The requirement of radiative electroweak symmetry breaking (REWSB)~\cite{Ibanez:1982fr} puts an important theoretical constraint on the parameter space. Another important constraint comes from limits on the cosmological abundance of stable charged particles~\cite{Yao:2006px}. This excludes regions in the parameter space where charged SUSY particles, such as $\ttau_1$ or $\tst_1$, become the lightest supersymmetric particle (LSP). We accept only those solutions for which the neutralino is the LSP.

We have performed random scans for the following parameter range:
\begin{eqnarray}
0\leq & m_{0} & \leq 20\, \rm{TeV}, \nonumber \\
0\leq & m_{1/2} & \leq 10\, \rm{TeV}, \nonumber  \\
-3 \leq & A_{0}/m_{0} & \leq 3 , \nonumber \\
5 \leq & \tan \beta & \leq 58,
\label{ppp1}
\end{eqnarray}
with $\mu >0~{\rm and}~ \mu<0$, and $m_t = 172.6$~GeV \cite{Group:2008nq}.

After collecting the data, we use the IsaTools package~\cite{Baer:2002fv} to implement the following phenomenological constraints:

$m_{\tilde{\chi}^{\pm}_{1}}~{\rm (chargino~mass)} \geq 103.5~{\rm GeV} \qquad   $ \hfill \cite{Yao:2006px},

$m_h~{\rm (lightest~Higgs~mass)} \geq 114.4~{\rm GeV} \qquad  $   \hfill  \cite{Schael:2006cr},

$BR(B_s \rightarrow \mu^+ \mu^-)< 5.8 \times 10^{-8} \qquad $  \hfill\cite{:2007kv},
 
$2.85 \times 10^{-4} \leq BR(B \rightarrow X_{s} \gamma)\leq 4.24 \times 10^{-4} \; (2\sigma) \; $ \hfill \cite{Barberio:2007cr}, 

$\Omega_{\rm CDM}h^2 = 0.111^{+0.011}_{-0.015} \;(2\sigma) \qquad \qquad $ \hfill \cite{Komatsu:2008hk},

$3.4 \times 10^{-10}\leq \Delta a_{\mu} \leq 55.6 \times 10^{-10}~ \; (3\sigma) \qquad $ \hfill \cite{Bennett:2006fi}.

We have applied the constraints from experimental data successively on the data that we acquired from ISAJET. As a first step we apply the constraints from $BR(B_s\rightarrow \mu^+ \mu^-)$, $BR(B\rightarrow X_{s} \gamma)$, chargino mass, and Higgs mass. We then apply the WMAP upper bound on the relic density of cold dark matter followed by the constraint on the muon anomalous magnetic moment $a_{\mu}=(g-2)_{\mu}/2$ at the $3\sigma$ allowed region. Finally, we apply the lower bound on the dark matter relic abundance. The data is then plotted showing the successive application of each of these constraints.

\section{Implications for CMSSM\label{results}}
PAMELA and ATIC, as discussed previously, have set constraints on the mass and other properties of the dark matter candidate which, in our CMSSM case, is the neutralino. The preferred mass values are as follows:

\begin{itemize}
\item  $m_{\tilde{\chi}^0_1} \gtrsim 0.3~{\rm TeV} $: This region is interesting if we wish to explain the PAMELA results based on decaying neutralinos.
\item $1.9~{\rm TeV} \lesssim m_{\tilde{\chi}^0_1} \lesssim 2.1~{\rm TeV}$: This relatively narrow range of neutralino mass explains the results from both PAMELA and ATIC based on decaying neutralinos.
\item $m_{\tilde{\chi}^0_1} \gtrsim 0.1~{\rm TeV}$: This region of neutralino masses can explain the PAMELA observation if we assume that neutralinos annihilate, rather than decay, preferentially into electrons and positrons. 
\item $0.65~{\rm TeV} \lesssim m_{\tilde{\chi}^0_1} \lesssim 0.75~{\rm TeV}$: This region can explain the results from both PAMELA and ATIC based on annihilating neutralinos.
\end{itemize}

We now present the results of the random scan focusing attention on the above mentioned neutralino mass ranges. In Fig.~\ref{fund1} we plot the results in the $(m_{1/2}, m_{0})$ and ($m_{1/2}$, $\tan\beta$) planes for  $\mu>0$ (left panel) and $\mu<0$ (right panel). All of these points satisfy the theoretical requirement of REWSB and correspond to a neutralino with mass less than $2.5$ TeV. In addition, these points satisfy the various experimental constraints listed earlier. Gray points satisfy the constraints from $BR(B_s\rightarrow \mu^+ \mu^-)$, $BR(B\rightarrow X_{s} \gamma)$, and the Higgs and chargino masses. After application of the WMAP 5 upper bound on neutralino dark matter (light blue points), the allowed region is drastically shrunk, with the remaining gray points associated with an unacceptably high dark matter relic density. The other colors (green, red, dark blue and black) in Fig.~\ref{fund1} correspond to different ranges of neutralino mass such that the neutralino satisfies both the WMAP 5 upper and lower bounds on dark matter relic abundance. The specific mass ranges have been picked out as interesting in trying to explain the data from PAMELA and ATIC. These are;
\begin{itemize}
\item  Green points correspond to $m_{\tilde{\chi}^0_1} \le 0.3~{\rm TeV} $.
\item  Black points correspond to $1.9~{\rm TeV} \le m_{\tilde{\chi}^0_1} \le 2.1~{\rm TeV}$.
\item  Red points correspond to $0.65~{\rm TeV} \le m_{\tilde{\chi}^0_1} \le 0.75~{\rm TeV}$.
\end{itemize}
Note that if we want to explain just the PAMELA data through neutralino annihilation with a suitable `boost' factor, there is almost no constraint on the CMSSM parameter space as it requires in this case $m_{\tilde{\chi}^0_1} \gtrsim 0.1~{\rm TeV}$, which is satisfied by almost all CMSSM allowed points.

The ($m_{1/2}$, $m_{0}$) plane for $\mu>0$ is distinctly different from that of $\mu<0$. With $\mu<0$, the ratio between $m_{1/2}$ and $m_{\tilde{\chi}^0_1}$ is fixed (more precisely in a narrow range), and the neutralino is always bino-like. While this is true also for the case $\mu>0$ for low values of $m_{1/2}$, we find solutions in which $m_{1/2}$ and $m_{\tilde{\chi}^0_1}$ lose this correlation for higher values, and the neutralino develops a strong higgsino component. This is the reason for the difference in the ($m_{1/2}$, $m_{0}$) planes between $\mu>0$ and $\mu<0$. This difference is more apparent in Fig.~\ref{spar2} where we plot results in the ($m_{\tilde g}$, $m_{\tilde{\chi}^0_1}$) plane.

In Fig.~\ref{fund2} we present the results in the ($m_{0}$, $A_{0}$) and ($m_{1/2}$, $A_0$) planes for both $\mu>0$ (left panel) and $\mu<0$ (right panel). Color coding is the same as in Fig. \ref{fund1}. Even though it is not explicitly shown in Fig.~\ref{fund2}, the constraint from $\Delta a_\mu$ favors a small magnitude ($\lesssim 4~{\rm TeV}$) for $A_0$ and $\mu>0$.

Let us now take a closer look at the sparticle spectroscopy that arises in the CMSSM. In Figs.~\ref{spar1}, \ref{spar2} and \ref{spar3} we present the results in the ($m_A$, $m_{\tilde{\chi}^0_1}$), ($m_{\tilde{\chi}^{\pm}_{1}}$, $m_{\tilde{\chi}^0_1}$), ($m_{\tilde g}$, $m_{\tilde{\chi}^0_1}$), ($m_{\tilde t}$, $m_{\tilde{\chi}^0_1}$), ($m_{h}$, $m_{\tilde{\chi}^0_1}$) and ($m_{\tilde \tau}$, $m_{\tilde{\chi}^0_1}$) planes for both $\mu>0$ (left panels) and $\mu<0$ (right panels). In these three figures we use consistent color coding. In particularl, points in gray satisfy the constraints from colliders on $BR(B_s\rightarrow \mu^+ \mu^-)$, $BR(B\rightarrow X_{s} \gamma)$, and the Higgs and chargino masses. Shown in light and dark blue are points that further satisfy bounds on the density of dark matter set by WMAP. Light blue points only satisfy the upper bound set by WMAP while dark blue ones satisfy both upper and lower bounds. Finally, we also show in orange and brown colors the regions allowed by the bounds on $\Delta a_\mu$. Orange points satisfy only the lower bound on dark matter relic density while brown points saturate the WMAP bound on the dark matter relic abundance.

The vertical lines in Figs.~\ref{spar1}, \ref{spar2} and \ref{spar3} correspond to $m_{\tilde{\chi}^0_1}=0.3,0.65,0.75,1.9~{\rm and}~2.1~{\rm TeV}$. These lines isolate the interesting ranges of neutralino mass from the point of view of PAMELA and ATIC as discussed above. 

Let us focus on trying to explain both PAMELA and ATIC using a decaying neutralino. This corresponds to the interval between the vertical lines $m_{\tilde{\chi}^0_1}=1.9~{\rm and}~2.1~{\rm TeV}$ in Figs.~\ref{spar1}, \ref{spar2} and \ref{spar3}. Of particular interest is the $\tilde t$ (stop) coannihilation region in Fig.~\ref{spar2}. In Fig.~\ref{spar3} we see that the stau is quite heavy for $1.9~{\rm TeV}\lesssim m_{\tilde{\chi}^0_1}\lesssim 2.1~{\rm TeV}$, so that coannihilation occurs mainly with the stop. The chargino can also be relatively `light' in these regions. However, it is clear from the ($m_{\tilde{\chi}^{\pm}_{1}}$, $m_{\tilde{\chi}^0_1}$) (Fig.~\ref{spar1}) and ($m_{\tilde t}$, $m_{\tilde{\chi}^0_1}$) (Fig.~\ref{spar2}) planes that the chargino and the stop are not simultaneously `light'. Therefore, to simultaneously explain both PAMELA and ATIC using a decaying neutralino requires a very heavy sparticle spectrum in the CMSSM, and the only particle that can be observed relatively easily at the LHC is the lightest Higgs boson.

If we focus our attention on explaining only the PAMELA results ($m_{\tilde{\chi}^0_1}\gtrsim 0.3~{\rm TeV}$), we obtain a sparticle spectrum that is much more accessible at the LHC. As shown in Fig.~\ref{spar1} in this case we can find a relatively light chargino ($\sim500{\rm GeV}$) and CP-odd and charged Higgs with masses $\sim600{\rm GeV}$. 

While we do not currently have a nice theoretical model for neutralino annihilations within the framework of CMSSM, it is plausible that such a model may be constructed in the future. If we allow for this possibility, it makes sense to explain ATIC and PAMELA in terms of neutralino annihilations and study the corresponding CMSSM sparticle spectroscopy. The advantage here is that we need a significantly lighter neutralino to explain the ATIC and PAMELA observations, and the corresponding spectroscopy is much more exciting from the point of view of the LHC.

In Table~\ref{table1} (Table~\ref{table2}) we present a few representative points for $\mu>0$ ($\mu<0$). All of these points correspond to a bino-like neutralino and have been chosen such that:
\begin{itemize}
\item Point 1 may be used to explain PAMELA based on either decaying or annihilating neutralinos.
\item Point 2 may be used to explain both ATIC and PAMELA based on annihilating neutralinos.
\item Point 3 may be used to explain both ATIC and PAMELA based on decaying neutralinos.
\end{itemize}
We also calculate for each benchmark point the neutralino-nucleon cross section. For $\mu>0$, the cross sections corresponding to the benchmark points 1 and 2 are a few orders of magnitude smaller then the exclusion limits set by the current experiments for direct dark matter detection such as CDMS \cite{CDMS} and XENON10 \cite{XENON}. The cross sections in the case of $\mu<0$ are smaller still. 

\section{Conclusions\label{conclusions}}

We have explored the implications for Higgs and sparticle spectroscopy of the CMSSM model in the light of recent cosmic ray observations reported by the  PAMELA and ATIC experiments. Our investigation is based on the premise that the lightest CMSSM neutralino comprises the dark matter in the universe, and that the underlying physics should provide a unified explanation, either through neutralino self annihilation or via its decay, for the positron and electron-positron excess reported by these two experiments. We have identified some benchmark points in the CMSSM parameter space which are consistent with this unified explanation. The corresponding sparticle mass spectra turns out, as expected, to be relatively heavy. The outlook for new particle discovery at the LHC is considerably enhanced if the ATIC data is simply ignored. In one example based on the decaying neutralino scenario, we find a relatively light chargino ($\sim500$ GeV) and CP-odd and charged Higgs  with masses $\sim600$ GeV. The CMSSM analysis reported here can be extended to other related scenarios with neutralino LSP such as those based on SU(5)~\cite{Profumo:2003ema}.

\acknowledgments
We thank Matt Kistler for helpful discussion and comments. This work is supported in part by the DOE Grant \# DE-FG02-91ER40626 (I.G., Q.S. and H.Y.),
GNSF grant 07\_462\_4-270 (I.G.), and by Bartol Research Institute (R.K.).
%


\end{document}